\documentstyle[amssymb,floats,aps,prb,epsf]{revtex}
\epsfclipon
\begin{document}
\twocolumn[\hsize\textwidth\columnwidth\hsize\csname
@twocolumnfalse\endcsname

\draft
\title{Indications of incommensurate spin fluctuations in doped
triangular antiferromagnets}
\author{Ying Liang}
\address{Department of Physics, Beijing Normal University, Beijing
100875, China}
\author{Shiping Feng}
\address{Department of Physics and Key Laboratory of Beam Technology
and Material Modification, Beijing Normal University, Beijing
100875, China\\
National Laboratory of Superconductivity, Academia Sinica, Beijing
100080, China}
\maketitle
%\date{\today}
\begin{abstract}
The incommensurate spin fluctuation of the doped triangular
antiferromagnet is studied within the $t$-$J$ model. It is shown
that the commensurate peak near the half-filling is split into six
incommensurate peaks in the underdoped and optimally doped regimes.
The incommensurability increases with the hole concentration at
lower dopings, and saturates at higher dopings. Although the
incommensurability is almost energy independent, the weight of
these incommensurate peaks decreases with energy and temperature.
\end{abstract}
]

\bigskip
\narrowtext

The high-$T_{c}$ cuprate superconductors exhibit many unusual
properties \cite{n1,n2}, one of the most striking being the
anomalous incommensurate antiferromagnetism in the normal state in
the underdoped regime \cite{n3,n4}. These unusual properties are
closely related to the fact that cuprate superconductors are doped
Mott insulators, obtained by chemically adding charge carriers to
a strongly correlated antiferromagnetic (AF) insulating state
\cite{n1,n2}. Neutron-scattering measurements show that when the
commensurate AF long-range-order (AFLRO) phase is suppressed and
the hole concentration exceeds $3\%$, incommensurate dynamical
short-range magnetic correlations appear \cite{n3,n4}, with four
peaks located at the reciprocal space positions $[(1\pm
\delta)\pi,\pi]$ and $[\pi,(1\pm\delta)\pi]$ (square lattice
notations, unit lattice constant). Even more remarkable is that
for very low dopings the incommensurability $\delta$ varies almost
linearly with the concentration $x$, but saturates at higher
dopings. Moreover, the incommensurate peaks broaden and weaken in
amplitude as the energy increases. This incommensurate
antiferromagnetism of doped cuprates results from special
microscopic conditions \cite{n3,n4}: (1) Cu ions situated in a
square-planar arrangement and bridged by oxygen ions, (2) weak
coupling between neighboring layers, and (3) doping in such a way
that the Fermi level lies near the middle of the Cu-O $\sigma^{*}$
bond. One common feature of these doped cuprates is the {\it
square-planar} Cu arrangement. However, some materials with a
two-dimensional (2D) spin arrangements on non-square lattices have
been synthetized. In particular, it has been reported \cite{n5}
that there is a class of doped cuprates, RCuO$_{2+\delta}$, R
being a rare-earth element, where the Cu ions sites sit not on a
square-planar, but on a {\it triangular-planar lattice}, therefore
allowing a test of the geometry effect on the spin fluctuations,
while retaining some other unique microscopic features of the Cu-O
bond. In other words, the question is whether the incommensurate
magnetic fluctuations observed on the doped square antiferromagnet
exist also in the doped triangular antiferromagnet? In addition,
the doped triangular antiferromagnet, where geometric frustration
is present, is also of theoretical interest by himself, with many
unanswered fascinating theoretical questions \cite{n51}.
Historically the undoped triangular antiferromagnet was firstly
proposed to be a model for microscopic realization of the
resonating valence bond spin liquid due to the strong frustration
\cite{n52}. It has been argued \cite{n9} that this spin liquid
state plays a crucial role in the superconductivity of doped
cuprates. Moreover, it has been shown \cite{n54} that the doped
and undoped triangular antiferromagnets present a pairing
instability in an unconventional channel.

The incommensurate antiferromagnetism of the doped square
antiferromagnet has been extensively studied, and some
fundamentally different microscopic mechanisms have been
proposed \cite{n6,n7,n71} to explain the origin of the
incommensurate antiferromagnetism, where there is a general
consensus that incommensurate antiferromagnetism emerges
due to doped charge carriers. Recently, it has been shown very
clearly \cite{n8} that if the strong electron correlation is
treated properly and strong spinon-holon interaction is taken
into account, the $t$-$J$ model can correctly reproduce all
main features of the incommensurate antiferromagnetism in the
underdoped square antiferromagnets, including the doping
dependence of the incommensurate peak position and the energy
dependence of the amplitude of these peaks. Since the strong
electron correlations are present in both doped square and
triangular antiferromagnets, it is expected that the
unconventional incommensurate antiferromagnetism existing in
the doped square antiferromagnet may also be seen in the doped
triangular antiferromagnet. In this paper, the purpose is to
apply the successful approach \cite{n8} for the doped square
lattice antiferromagnet to triangular lattice. We hope that the
information from the present work may induce further experimental
works in doped antiferromagnets on the non-square lattice.
\begin{figure}[prb]
\epsfxsize=2.5in\centerline{\epsffile{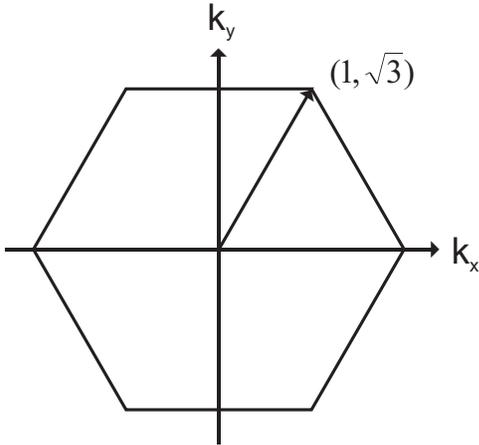}} \caption{The
Brillouin zone for the triangular system, where the
antiferromagnetic wave vector $Q=[1,\sqrt{3}]$.}
\end{figure}

As in the doped square antiferromagnet, the essential physics of
the doped triangular antiferromagnet is well described by the
$t$-$J$ model on the triangular lattice,
\begin{eqnarray}
H &=& -t\sum_{i\hat{\eta}\sigma}C^{\dagger}_{i\sigma}
C_{i+\hat{\eta}\sigma}+{\rm h.c.}-\mu \sum_{i\sigma}
C^{\dagger}_{i\sigma}C_{i\sigma} \nonumber \\
&+&J\sum_{i\hat{\eta}} {\bf S}_{i}\cdot {\bf S}_{i+\hat{\eta}},
\end{eqnarray}

with the electron single occupancy local constraint $\sum_{\sigma}
C^{\dagger}_{i\sigma}C_{i\sigma}\leq 1$, where the summation is
over all sites $i$, and for each $i$, over its nearest-neighbor
$\hat{\eta}$, ${\bf S}_{i}=C^{\dagger}_{i}{\vec \sigma}C_{i}/2$
are spin operators with ${\vec \sigma}=(\sigma_{x},\sigma_{y},
\sigma_{z})$ as Pauli matrices, and $\mu$ is the chemical
potential. In the $t$-$J$ model (1), the strong electron
correlation manifests itself by the local constraint \cite{n9},
therefore the crucial requirement is to impose this local
constraint. It has been shown \cite{n10} that this constraint can
be treated properly in analytical calculations within the
fermion-spin theory based on the charge-spin separation, where the
constrained electron operators are decomposed $C_{i\uparrow}=
h_{i}^{\dagger}S_{i}^{-}$ and $C_{i\downarrow}=h_{i}^{\dagger}
S_{i}^{+}$, where the spinless fermion operator $h_{i}$ describes
the charge (holon) degrees of freedom, while the pseudospin
operator $S_{i}$ describes the spin (spinon) degrees of freedom,
then the low energy behavior of the $t$-$J$ model (1) on the
triangular lattice in the fermion-spin representation can be
written as \cite{n10},
\begin{eqnarray}
H &=&-t\sum_{i\hat{\eta}}h_{i}h_{i+\hat{\eta}}^{\dagger}
(S_{i}^{+}S_{i+\hat{\eta}}^{-}+S_{i}^{-}S_{i+\hat{\eta}}^{+})
+\mu\sum_{i}h_{i}^{\dagger}h_{i} \nonumber \\
&+&J_{{\rm eff}}
\sum_{i\hat{\eta}}{\bf S}_{i}\cdot {\bf S}_{i+\hat{\eta}},
\end{eqnarray}
where $J_{{\rm eff}}=[(1-\delta)^{2}-\phi^{2}]$, $\delta$ is the
hole doping concentration, and $\phi=\langle h_{i}^{\dagger}
h_{i+\hat{\eta}}\rangle$ is the holon particle-hole order
parameter. At half-filling, the $t$-$J$ model is simply the
Heisenberg model. Many authors have shown that as in the square
lattice, there is indeed AFLRO in the ground state of the AF
triangular Heisenberg model \cite{n11}, but this AFLRO is
destroyed more rapidly with increasing dopings than on the
square lattice due to the strong geometric frustration, then away
from the half-filling, there is no AFLRO for the doped triangular
antiferromagnet, {\it i.e.}, $\langle S_{i}^{z}\rangle =0$, and
holons move self-consistently in the background of the spinon
liquid state.
\begin{figure}[prb]
\epsfxsize=2.5in\centerline{\epsffile{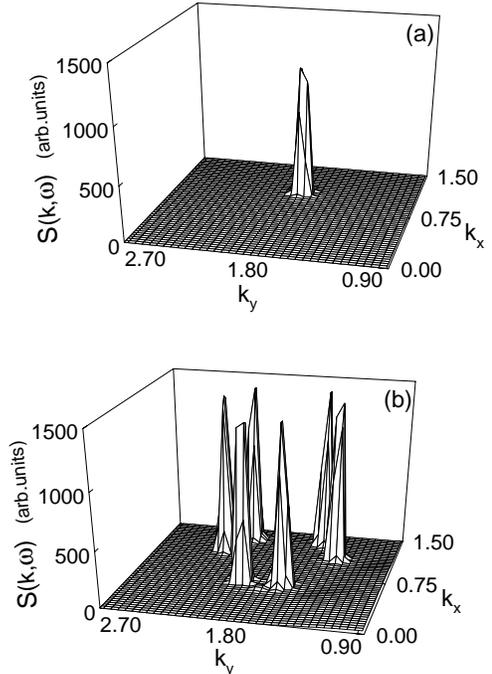}} \caption{The
dynamical spin structure factor in the $(k_{x},k_{y})$ plane at
doping (a) $x=0.015$ and (b) $x=0.06$ at temperature $T=0.15J$ and
energy $\omega=0.05J$ for parameter $t/J=2.5$.}
\end{figure}

Within the framework of the fermion-spin theory, the spin
fluctuations couple only to spinons, but the strong correlation
between holons and spinon is included through the holon's order
parameter $\phi$ entering the spinon propagator, therefore both
spinons and holons are responsible for the spin fluctuation. In
this case, the incommensurate spin dynamics in the doped square
antiferromagnet and integrated spin response in the doped
triangular antiferromagnet have been discussed \cite{n8,n12} by
considering the spin fluctuation around the mean-field (MF)
solution, where the spinon part is treated by a loop expansion
to second-order. Following their discussions \cite{n8,n12},
we can obtain the dynamical spin structure factor for the doped
triangular antiferromagnet as,
\begin{eqnarray}
S({\bf k},\omega)=-2[1+n_{B}(\omega)]{\rm Im}D({\bf k},\omega)
=-2[1+n_{B}(\omega)]\nonumber \\
\times{B^{2}_{k}{\rm Im}\Sigma_{s}({\bf k},\omega) \over
[\omega^{2}-\omega^{2}_{k}-B_{k}{\rm Re}\Sigma_{s}({\bf k},
\omega)]^{2}+[B_{k}{\rm Im}\Sigma_{s}({\bf k},\omega)]^{2}},
\end{eqnarray}
where the full spinon Green's function, $D^{-1}({\bf k},\omega)=
D^{(0)-1}({\bf k},\omega)-\Sigma_{s}({\bf k},\omega)$, and the MF
spinon Green's function $D^{(0)-1}({\bf k},\omega)=(\omega^{2}-
\omega_{k}^{2})/B_{k}$, ${\rm Im}\Sigma_{s}({\bf k},\omega)$
and ${\rm Re}\Sigma_{s}({\bf k},\omega)$ being the imaginary and
real parts of the second order spinon self-energy, respectively.
They are obtained from the holon bubble,
\begin{eqnarray}
\Sigma_{s}({\bf k},\omega)&=&-\left ({Zt\over N}\right )^{2}
\sum_{pp'}(\gamma_{k-p}+\gamma_{p'+p+k})^{2}{B_{k+p'}\over
2\omega_{k+p'}} \nonumber \\
&\times & \left ({F_{1}(k,p,p')\over \omega +\xi_{p+p'}-
\xi_{p}+\omega_{k+p'}} \right.\nonumber \\
&-&\left.{F_{2}(k,p,p')\over \omega
+\xi_{p+p'}-\xi_{p}-\omega_{k+p'}}\right ),
\end{eqnarray}
where $B_{k}=\Delta [(2\epsilon\chi_{z}+\chi)\gamma_{k}-(\epsilon
\chi+2\chi_{z})]$, $\Delta=2ZJ_{eff}$, $\epsilon=1+2t\phi/J_{eff}$,
$\gamma_{k}=[\cos{k_{x}}+2\cos{(k_{x}/2)}\cos{({\sqrt 3}k_{y}/2)}]
/3$, $Z$ is the number of the nearest neighbor sites,
$F_{1}(k,p,p')=n_{F}(\xi_{p+p'})[1-n_{F}(\xi_{p})]+[1+n_{B}
(\omega_{k+p'})][n_{F}(\xi_{p})-n_{F}(\xi_{p+p'})]$, $F_{2}
(k,p,p')=n_{F}(\xi_{p+p'})[1-n_{F}(\xi_{p})]-n_{B}(\omega_{k+p'})
[n_{F}(\xi_{p})-n_{F}(\xi_{p+p'})]$, $n_{F}(\xi_{k})$ and $n_{B}
(\omega_{k})$ are the fermion and boson distribution functions,
respectively, the MF holon excitation $\xi_{k}=2\chi tZ\gamma_{k}
+\mu$, and MF spinon excitation, $\omega^{2}_{k}=\Delta^{2}(A_{1}
\gamma^{2}_{k}+A_{2}\gamma_{k}+A_{3})$ with $A_{1}=\alpha\epsilon
(\chi/2+\epsilon\chi_{z})$, $A_{2}=\epsilon[(1-Z)\alpha(\epsilon
\chi/2+\chi_{z})/Z-\alpha (C_{z}+C/2)-(1-\alpha)/(2Z)]$, $A_{3}=
\alpha(C_{z}+\epsilon^{2} C/2)+(1-\alpha)(1+\epsilon^{2})/(4Z)-
\alpha\epsilon(\chi/2+\epsilon\chi_{z})/Z$, the spinon correlation
functions $\chi=\langle S_{i}^{+}S_{i+\hat{\eta}}^{-}\rangle$,
$\chi_{z}=\langle S_{i}^{z}S_{i+\hat{\eta}}^{z}\rangle$,
$C=(1/Z^{2})\sum_{\hat{\eta}\hat{\eta'}}\langle
S_{i+\hat{\eta}}^{+}S_{i+\hat{\eta'}}^{-}\rangle$, and $C_{z}=
(1/Z^{2})\sum_{\hat{\eta}\hat{\eta'}}\langle S_{i+\hat{\eta}}^{z}
S_{i+\hat{\eta'}}^{z}\rangle$. In order to satisfy the sum rule
for the correlation function $\langle S^{+}_{i}S^{-}_{i}
\rangle=1/2$ in the absence of AFLRO, a decoupling parameter
$\alpha$ has been introduced in the self-consistent MF
calculation, which can be regarded as the vertex correction
\cite{n13}.
\begin{figure}[prb]
\epsfxsize=2.5in\centerline{\epsffile{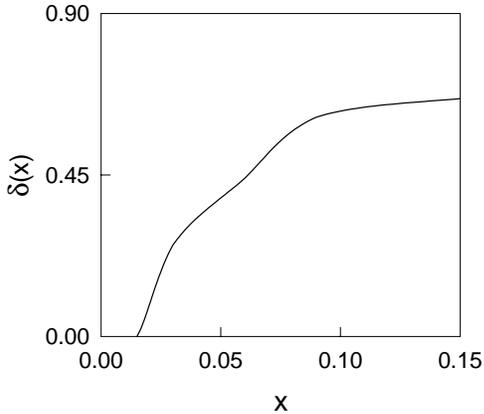}} \caption{The
doping dependence of the incommensurability $\delta(x)$ of the
antiferromagnetic fluctuations.}
\end{figure}

The Brillouin zone of the triangular system is shown in Fig. 1,
where the AF wave vector ${\bf Q}=[1,\sqrt{3}]$ (hereafter we use
the units of $[2\pi/3,2\pi/3]$). In Fig. 2, we plot the dynamical
spin structure factor $S({\bf k},\omega)$ in the [$k_{x},k_{y}$]
plane at doping (a) $x=0.015$ and (b) $x=0.06$ with temperature
$T=0.15J$ and energy $\omega=0.05J$ for parameter $t/J=2.5$. From
Fig. 2, we find that a commensurate-incommensurate transition in
the spin fluctuation ${\bf Q}$ vector occurs with doping, {\it
i.e.}, the commensurate peak near half-filling ($x\leq 0.015$) is
split into six peaks in the underdoped regime, while the positions
of these split peaks are incommensurate with the underlying
lattice, and correspond to six 2D rods at
$[(1-\delta_{x}),(\sqrt{3}\pm\delta_{y})]$, $[(1-\delta'_{x}),
(\sqrt{3}\pm \delta'_{y})]$, and $[(1-\delta''_{x}),(\sqrt{3} \pm
\delta''_{y})]$ with $\sqrt{\delta^{2}_{x}+\delta^{2}_{y}}=
\sqrt{(\delta'_{x})^{2}+(\delta'_{y})^{2}}=
\sqrt{(\delta''_{x})^{2}+(\delta''_{y})^{2}}=\delta$. At low
temperatures, the incommensurate peaks are very sharp at lower
energies, which means that these excitations have a dynamical
coherence length that is larger than the instantaneous correlation
length. The present dynamical spin structure factor spectrum
$S({\bf k},\omega)$ has been used to extract the doping dependence
of the incommensurability parameter $\delta(x)$, defined as the
deviation of the peak position from the AF wave vector ${\bf Q}$,
and the result is plotted in Fig. 3. Our result shows that
$\delta(x)$ increases with the hole concentration in lower
dopings, but it saturates at higher dopings, which is
qualitatively similar to the results for the square
antiferromagnet \cite{n3,n4,n8}.
\begin{figure}[prb]
\epsfxsize=2.5in\centerline{\epsffile{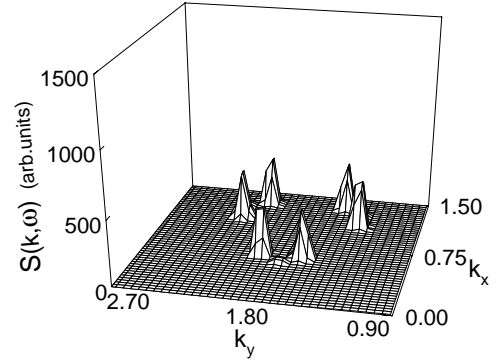}} \caption{The
dynamical spin structure factor in the $(k_{x},k_{y})$ plane at
doping $x=0.06$ for parameter $t/J=2.5$ and energy $\omega=0.1J$
at temperature $T=0.15J$.}
\end{figure}
\begin{figure}[prb]
\epsfxsize=2.5in\centerline{\epsffile{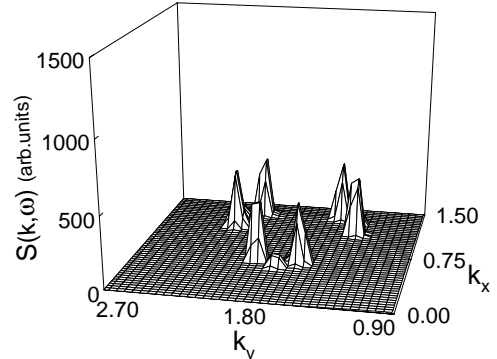}} \caption{The
dynamical spin structure factor in the $(k_{x},k_{y})$ plane at
doping $x=0.06$ for parameter $t/J=2.5$ and energy $\omega=0.05J$
at temperature $T=0.3J$.}
\end{figure}

In order to study the effects of energy and temperature on the
incommensurate spin fluctuation, we have calculated the dynamical
spin structure factor $S({\bf k},\omega)$ at different
temperatures and energies, and the result at doping $x=0.06$ for
parameter $t/J=2.5$, temperature $T=0.15J$ and energy
$\omega=0.1J$ is shown in Fig. 4. In comparison with Fig. 2(b) for
the same set of parameters except for $\omega=0.05J$, we see that
although the positions of the incommensurate peaks are almost
energy independent, the weight of these peaks decreases with
increasing energy, and tends to vanish at high energies. This
reflects that the inverse lifetime of the spin excitations
increases with increasing energy. In correspondence, the dynamical
spin structure factor $S({\bf k},\omega)$ at doping $x=0.06$ for
parameter $t/J=2.5$ in energy $\omega=0.05J$ and temperature
$T=0.3J$ is plotted in Fig. 5. Comparing with Fig. 2(b) for the
same set of parameters except for $T=0.15J$, we find that the peak
weight is suppressed with increasing temperature. Our results also
indicate that although the positions of the incommensurate peaks
in the doped triangular antiferromagnets are much different from
these in the doped square antiferromagnets due to the geometric
frustration, the behavior of the energy and temperature dependence
of the magnetic fluctuation is qualitatively similar to these in
the doped square antiferromagnets \cite{n3,n4,n8}.
\begin{figure}[prb]
\epsfxsize=2.5in\centerline{\epsffile{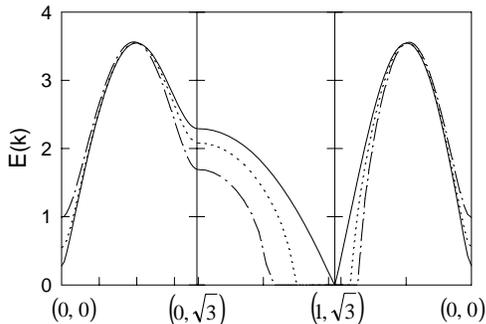}} \caption{The
spinon excitation spectrum at doping $x=0.015$ (solid line),
$x=0.03$ (dashed line), and $x=0.06$ (dot-dashed line) in
parameter $t/J=2.5$ with temperature $T=0.15J$.}
\end{figure}
\begin{figure}[prb]
\epsfxsize=2.5in\centerline{\epsffile{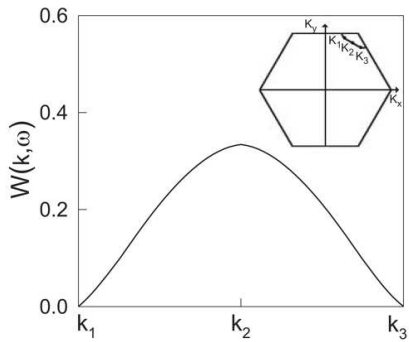}} \caption{Function
$W({\bf k},\omega)$ at doping $x=0.06$ for parameter $t/J=2.5$ and
energy $\omega=0.05J$ at temperatures $T=0.15J$.}
\end{figure}

As in the case of the doped square antiferromagnet \cite{n8},
the effect of holons on the spinons due to the strong
spinon-holon interaction is critical in determining the
characteristic feature of the incommensurate spin fluctuation
in the doped triangular antiferromagnet. This is reflected by
the renormalization of the spinon excitation
$E^{2}_{k}=\omega^2_{k}+B_{k}{\rm Re}\Sigma_{s}(k,E_{k})$ in
Eq. (3). Figure 6 shows this renormalized spinon excitation for
doping $x=0.015$ (solid line), $x=0.03$ (dashed line), and
$x=0.06$ (dot-dashed line), and for $t/J=2.5$ and $T=0.15J$. An
over damping around the AF wave vector ${\bf Q}$ occurs for doping
$x\geq 0.015$. The incommensurate peaks emerge in the dynamical
spin structure factor $S({\bf k},\omega)$ when the incoming
neutron energy $\omega$ is equal to the renormalized spin
excitation $E_{k}$, {\it i.e.} $W({\bf k}_{\delta},\omega)\equiv
[\omega^{2}-\omega^{2}_{k_{\delta}}-B_{k_{\delta}}{\rm Re}
\Sigma_{s}({\bf k}_{\delta},\omega)]^{2}=(\omega^{2}-
E^{2}_{k_{\delta}})^{2}\sim 0$ for some critical wave vectors
${\bf k}_{\delta}$ (positions of the incommensurate peaks). Then
the weight of these peaks is determined by the imaginary part of
the spinon self-energy $1/{\rm Im}\Sigma_{s}({\bf k}_{\delta},
\omega)$. Therefore near half-filling, the spin excitations are
centered around the AF wave vector ${\bf Q}$, so the commensurate
AF peak appears there. Upon doping, the holes disturb the AF
background self-consistently, and induce over damping of the
spinon excitation around the AF wave vector ${\bf Q}$, leading to
incommensurate antiferromagnetism. Since the weight of the
incommensurate peaks is determined by the imaginary part of the
spinon self-energy, it is then understandable that they are
suppressed as energy and temperature are increased. Moreover, the
symmetry of the incommensurate pattern is determined by the
lattice symmetry in the spinon self-energy renormalization due to
holons. We have plotted the function $W({\bf k},\omega)$ along the
line shown in the inset of Fig. 7 at doping $x=0.06$ for parameter
$t/J=2.5$ and temperature $T=0.15J$ and energy $\omega=0.05J$ in
Fig. 7. There is a strong angular dependence with minima at
$[(1-\delta_{x}),(\sqrt{3}-\delta_{y})]$, $[(1-\delta'_{x}),
(\sqrt{3}-\delta'_{y})]$. These are exactly the positions of the
incommensurate peaks determined by the dispersion of very well
defined renormalized spin excitations. This incommensurate pattern
is much different from that in the doped square antiferromagnet
\cite{n8}, since the lattice symmetry in the triangular lattice
is much different from that in the square lattice.

In summary, we have discussed the incommensurate spin fluctuation
of the doped triangular antiferromagnet within the $t$-$J$ model
based on the fermion-spin theory. It is shown that the
commensurate peak near the half-filling is split into six
incommensurate peaks in the underdoped and optimally doped regimes.
The incommensurability increases with the hole concentration at
lower dopings, and saturates at higher dopings. Although the
incommensurability is almost energy independent, the weight of
these incommensurate peaks decreases with energy and temperature.

\acknowledgments
The authors would like to thank Dr. Feng Yuan for helpful
discussions. This work was supported by the National Natural
Science Foundation under the Grant No. 10074007, 10125415, and
90103024, and the Grant from Ministry of Education of China.

\end{document}